\documentclass[aps,prd,twocolumn,floatfix,nofootinbib]{revtex4-1}

\usepackage{graphicx}
\usepackage{amsmath}
\usepackage{amssymb}
\usepackage{xcolor}
\usepackage[colorlinks,citecolor=blue,linkcolor=blue,urlcolor=blue,anchorcolor=blue]{hyperref}
\allowdisplaybreaks
\newcommand\dx{{\rm d}}
\newcommand\p{\partial}
\newcommand\etal{{\it et~al.}}

\newcommand\aj{Astron. J.}
\renewcommand\apj{Astrophys. J.}
\newcommand\apjl{Astrophys. J. Lett.}

\newcommand\epjc{Eur. Phys. J. C}

\newcommand\jcap{J. Cosmol. Astropart. Phys.}

\newcommand\mnras{Mon. Not. R. Astron. Soc.}
\newcommand\mpla{Mod. Phys. Lett. A}

\newcommand\pdu{Phys. Dark Universe}
\newcommand\plb{Phys. Lett. B}
\newcommand\prep{Phys. Rep.}

\begin{document}

\title{Cosmological consequences of a scalar field with oscillating equation of state:
  A possible solution to the fine-tuning and coincidence problems}
\author{S. X. Ti\'an}
\email[]{tshuxun@whu.edu.cn}
\affiliation{School of Physics and Technology, Wuhan University, 430072 Wuhan, China}
\date{\today}
\begin{abstract}
  We propose a new dark energy model for solving the cosmological fine-tuning and coincidence problems. A default assumption is that the fine-tuning problem disappears if we do not interpret dark energy as vacuum energy. The key idea to solving the coincidence problem is that the Universe may have several acceleration phases across the whole cosmic history. The specific example we study is a quintessence model with approximately repeated double exponential potential, which only introduces one Planck scale parameter and three dimensionless parameters of order unity. The cosmological background evolution equations can be recast into a four-dimensional dynamical system and its main properties are discussed in details. Preliminary calculations show that our model is able to explain the observed cosmic late-time acceleration.
\end{abstract}
\maketitle

\section{Introduction}\label{sec:01}
The cosmic late-time acceleration has been confirmed for two decades \cite{Riess1998,Perlmutter1999}. The standard model to explain this phenomenon is the $\Lambda$CDM model with the cosmological constant $\Lambda=\mathcal{O}(H_0^2/c^2)$, where $H_0$ is the Hubble constant. This model suffers from two tricky problems: the fine-tuning and coincidence problems. The energy density of vacuum given by quantum field theory should be of the order of Planck scale density $\rho_{\rm P}=5.1\times10^{96}\ {\rm kg}/{\rm m}^3$ while the observed effective energy density of $\Lambda$ is $\rho_\Lambda=6.6\times10^{-27}\ {\rm kg}/{\rm m}^3$. The ratio of $\rho_\Lambda$ and $\rho_{\rm P}$ is $\mathcal{O}(10^{-120})$. If one interprets the origin of $\Lambda$ as the vacuum energy, then how to obtain $\rho_\Lambda$ from $\rho_{\rm P}$ is a fine-tuning problem. There are many attempts to explain the origin of this extremely small ratio, e.g., spacetime foam \cite{Wang2017,Wang2019a,Wang2019b} and quantum gravity discreteness \cite{Josset2017,Perez2019}. However, a mature theory in this way seems far away from us \cite{Bengochea2020}. If we abandon  $\Lambda$ but use a dynamical field to explain the late-time acceleration, then we can completely hide the vacuum energy at macroscopic scales (i.e., solve the old cosmological constant problem) with reasonable theories \cite{Weinberg1989,Carlip2019}. The popular dynamical fields that used to explain the late-time acceleration include quintessence \cite{Caldwell1998,Steinhardt1999,Zlatev1999}, phantom \cite{Caldwell2002}, quintom \cite{Feng2005,Guo2005,Zhao2006a,Cai2010}, etc. However, the coincidence problem still exists in these dynamical dark energy models. In addition to the dark energy model, modified gravity is also widely used to explain the cosmic late-time acceleration \cite{Clifton2012}. However, none of these gravity theories can naturally solve the problems we are considering (see discussions in Ref. \cite{Bull2016} and references therein).

The coincidence problem  states why the effective dark energy density $\rho_{\rm DE}$ is comparable to the normal matter density $\rho_{\rm m}$ at today. Note that it is controversial that whether the cosmic coincidence is problematic \cite{Bianchi2010,Sivanandam2013,Velten2014}. If the cosmic expansion is parameterized in terms of the redshift or logarithmic scale factor, then the coincidence problem do exist. But the problem disappears if we use the cosmic time or linear scale factor to parameterize the expansion. We think the most natural axis is the logarithmic scale factor because it characterizes the order of magnitude of $\rho_{\rm m}$. In this paper, we give tacit consent to existence of the coincidence problem described with the logarithmic scale factor. This is subjective. Anyway, the solution to the possible coincidence problem adds new motivation to the theory. It is believed that some dynamical dark energy models can alleviate, but not solve, the coincidence problem with the tracker property \cite{Peebles1988,Ratra1988,Steinhardt1999,Zlatev1999,Zhao2006b}. However, we may rephrase the coincidence problem as why the transition from matter-dominated Universe to dark energy-dominated Universe occurs today. In this sense, the tracker property does nothing to alleviate the coincidence problem. Intuitively, the coincidence problem disappears if the Universe was dominated by normal matters and dark energy alternately across the whole cosmic history (see $\Omega_\phi$ in Fig. \ref{fig:02} for an intuition). In this scenario, $\rho_{\rm DE}$ tracks $\rho_{\rm m}$ in the view of long time period. But the behaviors of $\rho_{\rm DE}$ and $\rho_{\rm m}$ are different in the view of short time period. Another property is that the corresponding dark energy equation of state (EoS) should be oscillating. There are many works to discuss the  oscillating dark energy EoS, e.g., Refs. \cite{Xia2005,Xia2006,Liu2009,Nojiri2006a,Nojiri2006b,Kurek2010,Lazkoz2010,Pan2018,Panotopoulos2018, Brownsberger2019,Rezaei2019,Tamayo2019}. However, no model meets our requirements (see Appendix \ref{sec:A}).

In addition, we require that human-related parameters should not be introduced into the fundamental theory. The wide-used dark energy models \cite{Caldwell1998,Steinhardt1999,Zlatev1999,Caldwell2002,Feng2005,Guo2005,Zhao2006a,Cai2010} generally need a parameter related to $H_0$. The unnatural point is that $H_0$ is the value of the Hubble parameter at today (the time human exist). Note that, as we discussed before, in this paper, we use the logarithmic scale factor to parameterize the cosmic expansion and thus $H_0$ is a special value. In principle, the fundamental physical constant that appears in cosmology theory can be any value less than Planck scale values. In other words, we may ask whether we can explain the observed cosmic late-time acceleration with a parameter of the order of $\mathcal{O}(10^9H_0)$ or $\mathcal{O}(10^{-9}H_0)$. Furthermore, we require the desired model only introduces Planck scale parameters and dimensionless parameters of order unity. This requirement is subjective but adds new motivation to the theory. Similar consideration was also discussed in Ref. \cite{Bordin2019}, which proposed a new dark energy model with $\mathcal{O}(100)$ scalar fields. Among the various modified gravities, nonlocal gravity provides the possibility to explain the late-time acceleration using only dimensionless parameters of order unity \cite{Deser2007,Tian2018}. However, observations about the gravitational bound systems rule out these theories \cite{Tian2019,Belgacem2019}.

How to construct a concrete dark energy model to realize the scenario described in the previous two paragraphs? Technically, in the framework of quintessence, we might be able to achieve this with a repeated double exponential potential (see Fig. \ref{fig:01} for an intuition). Note that the single steep exponential potential could realize $\rho_{\rm DE}$ tracks $\rho_{\rm m}$ \cite{Copeland1998}. The model with double exponential potential, in which one is steep and one is flat, presents both early scaling and late-time accelerating solutions \cite{Barreiro2000}. If we repeat the double exponential potential periodically, we may find a realization of the desired model. The specific model is described in Sec. \ref{sec:02} and its properties are analyzed in Sec. \ref{sec:03}.

\section{The model}\label{sec:02}
\begin{figure}[!b]
  \centering
  \includegraphics[width=0.99\linewidth]{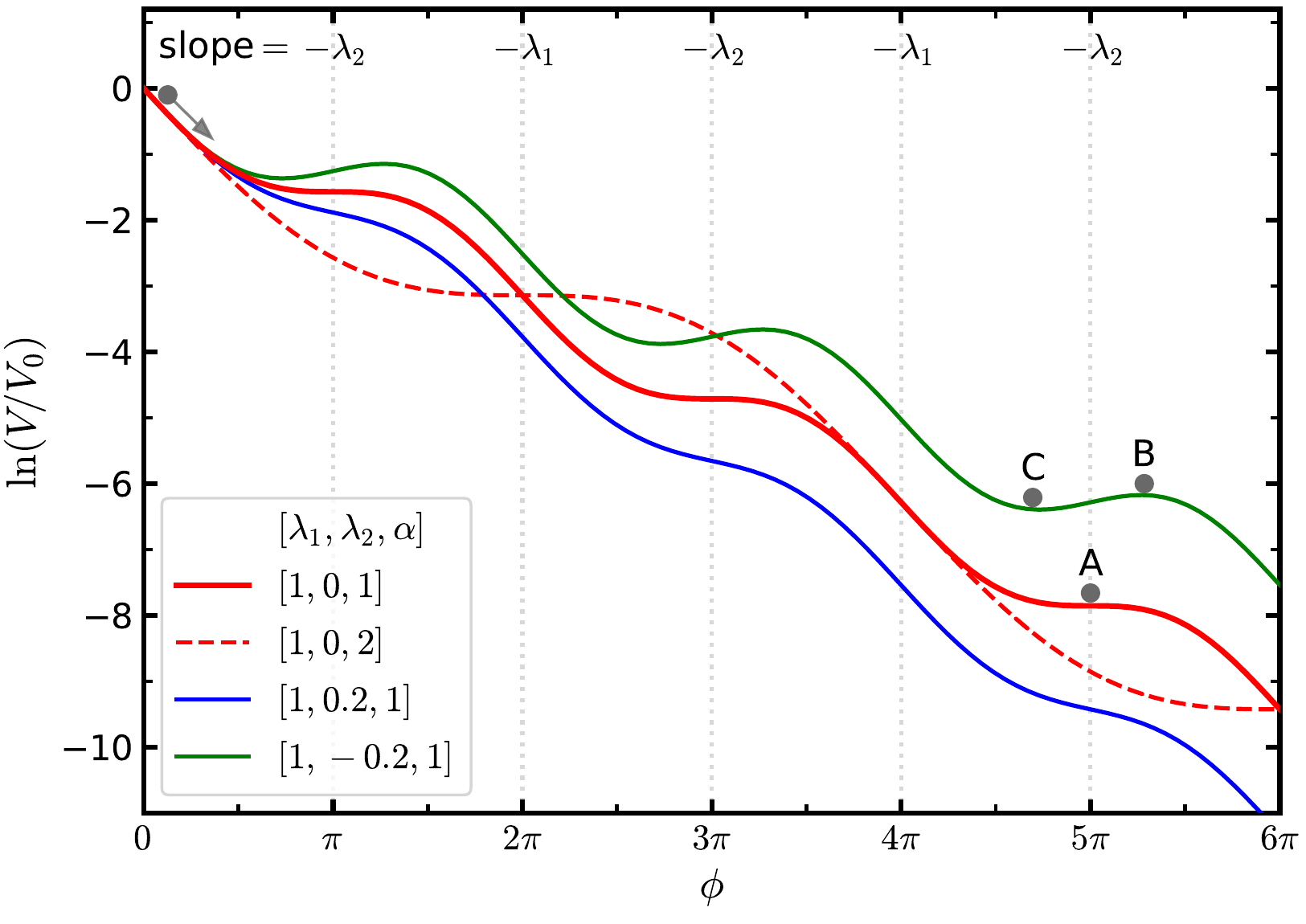}
  \caption{The potential of the sine oEoS model. The slope of the solid curve at $\phi=n\pi$ ($n$ is an integer) is written at the top of the figure. $V'=0$ at points A, B and C.}
  \label{fig:01}
\end{figure}

We consider cosmic expansion driven by a single scalar field with normal matters including radiation and dust. The action for this physical system is of the form
\begin{equation}
  S=\int\dx^4x\sqrt{-g}\left[\frac{R}{2\kappa}-\frac{\mathcal{L}_\phi}{\kappa}\right]+S_\textrm{m},
\end{equation}
where $\kappa\equiv8\pi G/c^4$. For the normal matters, we know the variation $\delta S_{\rm m}=-\frac{1}{2}\int\dx^4x\sqrt{-g}T_{\mu\nu}\delta g^{\mu\nu}$ and $T_{\mu\nu}=\left(\rho_{\rm m}+p_{\rm m}/c^2\right)u_\mu u_\nu+p_{\rm m}g_{\mu\nu}$. The EoS of normal matters is defined as $w_{\rm m}\equiv p_{\rm m}/(\rho_{\rm m}c^2)$. We know $w_{\rm m}=0$ for the dust and $w_{\rm m}=1/3$ for the radiation. For the scalar field, we adopt $\mathcal{L}_\phi=X+V(\phi)$, where $X=\frac{1}{2}g^{\mu\nu}\partial_\mu\phi\partial_\nu\phi$. For the potential, we do not repeat the double exponential potential exactly, but assume
\begin{equation}\label{eq:02}
  V(\phi)=V_0\exp\left[-\frac{\lambda_1+\lambda_2}{2}\phi
  -\frac{\alpha(\lambda_1-\lambda_2)}{2}\sin\frac{\phi}{\alpha}\right],
\end{equation}
where $V_0$, $\lambda_i$ and $\alpha$ are parameters. In our conventions, $\phi$, $\lambda_i$ and $\alpha$ are dimensionless and $[V_0]={\rm length}^{-2}$. Equation (\ref{eq:05a}) requires $V_0>0$. For the first step, we can assume $\lambda_1>0$, $|\lambda_2|<\lambda_1$ and $\alpha>0$. Figure \ref{fig:01} plots $V(\phi)$ for four cases. The $\alpha$ controls the period of oscillation. The $\lambda$ varies as $\lambda_1\rightarrow\lambda_2\rightarrow\lambda_1\rightarrow\lambda_2\rightarrow\cdots$ with $\phi$ increasing [see Eq. (\ref{eq:07b}) for the definition of $\lambda$]. This potential can be regarded as an approximate but simple realization of the repeated double exponential potential. For suitable parameter settings, we expect the Universe is decelerating when $\lambda\approx\lambda_1$ and accelerating when $\lambda\approx\lambda_2$. However, this is just our initial idea. The system behaves much more complex as we will see in Sec. \ref{sec:03}. Variation of the action with respect to the metric gives the gravitational field equations
\begin{equation}
  G_{\mu\nu}=\kappa T_{\mu\nu}+\Phi_{\mu\nu},
\end{equation}
where $\Phi_{\mu\nu}=\p_\mu\phi\p_\nu\phi-g_{\mu\nu}\mathcal{L}_\phi$. Variation of the action with respect to $\phi$ gives the scalar field equation $\Box\phi=V'$, where $V'\equiv\dx V/\dx\phi$. Hereafter we call the model described by Eq. (\ref{eq:02}) as the sine oEoS model, where the first letter o means oscillating and EoS means the equation of state. Replacing the sine with the cosine in Eq. (\ref{eq:02}) does not change the essence of the model as $\cos(x)=\sin(x+\pi/2)$.

To be consistent with current observations \cite{Aghanim2018}, we assume the Universe is described by the flat Friedmann-Lema\^{i}tre-Robertson-Walker metric
\begin{equation}
  \dx s^2=-c^2\dx t^2+a^2(\dx x^2+\dx y^2+\dx z^2),
\end{equation}
where $a=a(t)$. For the normal matters, the energy-momentum tensor is $T^\mu_{\ \nu}=\textrm{diag}\{-\rho_{\rm m}c^2,p_{\rm m},p_{\rm m},p_{\rm m}\}$ and energy conservation is described by $\dot{\rho}_{\rm m}+3(1+w_{\rm m})H\rho_{\rm m}=0$. For the scalar field, we can assume $\phi=\phi(t)$, which gives $X=-\dot{\phi}^2/(2c^2)$ and  $\Phi^\mu_{\ \nu}={\rm diag}\{X-V,-X-V,-X-V,-X-V\}$. Substituting the above results into the gravitational and scalar field equations, we obtain
\begin{subequations}\label{eq:05}
\begin{gather}
  H^2=\frac{8\pi G}{3}\rho_{\rm tot},\label{eq:05a}\\
  \frac{\ddot{a}}{a}=-\frac{4\pi G}{3}(\rho_{\rm tot}+\frac{3p_{\rm tot}}{c^2}),\label{eq:05b}\\
  \ddot{\phi}+3H\dot{\phi}+c^2V'=0,\label{eq:05c}
\end{gather}
\end{subequations}
where $\rho_{\rm tot}=\rho_{\rm m}+\rho_\phi$, $p_{\rm tot}=p_{\rm m}+p_\phi$, $\rho_\phi\equiv(-X+V)/(\kappa c^2)$ and $p_\phi\equiv(-X-V)/\kappa$. Equivalently, we can define the EoS of the scalar field as $w_\phi\equiv p_\phi/(\rho_\phi c^2)$ and Eq. (\ref{eq:05c}) can be written as $\dot{\rho}_\phi+3(1+w_\phi)H\rho_\phi=0$. Equation (\ref{eq:05c}) can be derived from Eqs. (\ref{eq:05a}) and (\ref{eq:05b}) as we expected. In order to compare theoretical results with observations, we define
\begin{gather}
  \Omega_\phi\equiv\frac{8\pi G}{3H^2}\rho_\phi,\
  \Omega_{\rm m}\equiv\frac{8\pi G}{3H^2}\rho_{\rm m}=1-\Omega_\phi,\nonumber\\
  w_{\rm tot}\equiv\frac{p_{\rm tot}}{\rho_{\rm tot}c^2}=\Omega_{\rm m}w_{\rm m}+\Omega_\phi w_\phi.\label{eq:06}
\end{gather}
All the fitting parameters of the cosmological constraints about the cosmic background evolution are enclosed in $\{w_{\rm tot},H_0\}$. For example, the luminosity distance $D_L(z)=\frac{(1+z)c}{H_0}\int_0^z\frac{\dx\tilde{z}}{E(\tilde{z})}$, where $E^2(z)=\exp(\int_0^z\frac{3[1+w_{\rm tot}(\tilde{z})]}{1+\tilde{z}}\dx\tilde{z})$. In this paper, we do not fit real data because the chaos phenomenon in the model makes the classical statistical methods invalid (see Sec. \ref{sec:03} for detailed discussions). Instead, in the next section, we discuss how well the sine oEoS model with certain parameter settings can recover the $\Lambda$CDM model in the late-time era.

\section{Main properties}\label{sec:03}
Phase space analysis is a powerful tool for quantitatively understanding the cosmological dynamics \cite{Bahamonde2018}. As in the case of exponential potential \cite{Copeland1998}, we define the dimensionless variables
\begin{subequations}\label{eq:07}
\begin{equation}
  x_1\equiv\frac{\dot{\phi}}{\sqrt{6}H},\quad
  x_2\equiv\frac{c\sqrt{V}}{\sqrt{3}H}.
\end{equation}
As declared in the review paper \cite{Bahamonde2018}, these two variables were first introduced in Ref. \cite{Copeland1998}. Based on this definition, we have $\Omega_\phi=x_1^2+x_2^2$ and $w_\phi=(x_1^2-x_2^2)/\Omega_\phi$. For the sine oEoS model, we also need
\begin{align}
  \lambda&\equiv-\frac{V'}{V}=\frac{\lambda_1+\lambda_2}{2}+\frac{\lambda_1-\lambda_2}{2}\cos\frac{\phi}{\alpha},\label{eq:07b}\\
  \nu&\equiv\sqrt{6}(\lambda^2-\mu)=-\frac{\sqrt{6}(\lambda_1-\lambda_2)}{2\alpha}\sin\frac{\phi}{\alpha},
\end{align}
\end{subequations}
where $\mu\equiv V''/V$. The cosmic evolution equations can then be written as
\begin{subequations}\label{eq:08}
\begin{align}
  \frac{\dx x_1}{\dx N}&=-3x_1+\frac{\sqrt{6}}{2}\lambda x_2^2+\frac{3}{2}x_1L,\\
  \frac{\dx x_2}{\dx N}&=-\frac{\sqrt{6}}{2}\lambda x_1x_2+\frac{3}{2}x_2L,\\
  \frac{\dx\lambda}{\dx N}&=\nu x_1,\label{eq:08c}\\
  \frac{\dx\nu}{\dx N}&=\frac{3x_1}{\alpha^2}(\lambda_1+\lambda_2-2\lambda),
\end{align}
\end{subequations}
where $L=(1-w_{\rm m})x_1^2+(1+w_{\rm m})(1-x_2^2)$, $N\equiv\ln(a/a_i)$ and $a_i$ is the value of the scale factor at any fixed time point. Similar to the exponential and power-law potentials \cite{Copeland1998,Ng2001}, the parameter $V_0$ disappears in the dynamical system equations in our model. The evolution of this physical system is completely described by trajectories within the region $x_1^2+x_2^2\leqslant1$, $x_2\geqslant0$, $\lambda_2\leqslant\lambda\leqslant\lambda_1$ and $|\nu|\leqslant\sqrt{6}(\lambda_1-\lambda_2)/(2\alpha)$. One constraint equation given by Eq. (\ref{eq:07}) is
\begin{equation}\label{eq:09}
  \nu(\lambda)=\nu_\pm(\lambda)=\pm\frac{\sqrt{6}}{\alpha}\sqrt{\lambda(\lambda_1+\lambda_2)-\lambda^2-\lambda_1\lambda_2}.
\end{equation}
The sign of the above equation is changeable with the system evolution. This is why we do not substitute Eq. (\ref{eq:09}) into Eq. (\ref{eq:08c}) to eliminate $\nu$ to obtain a three-dimensional dynamical system. Using the four-dimensional dynamical system Eq. (\ref{eq:08}) to characterise the evolution of the Universe can avoid the sign selection problem, which is useful to the following analyses.

\subsection{Critical points and stability}
We are now ready to find the critical points of the four-dimensional dynamical system Eq. (\ref{eq:08}) and to perform the stability analysis. We only consider the case where $w_{\rm m}=0$ or $1/3$. Depending on the value of $\lambda_2$, we have up to four critical points which are listed in Table \ref{tab:01}. In principle, one can directly obtain the stability of points A, B and C from Fig. \ref{fig:01}. However we also perform the standard stability analysis (see Ref. \cite{Bahamonde2018} for a lecture).

\begin{table*}[!t]
\centering
\caption{Critical points of the dynamical system Eq. (\ref{eq:08}) with existence and physical properties. The label column is consistent with the labels in Fig. \ref{fig:01}. The methods to analyze the stability are listed in the last column, in which linear stability theory is performed to Eq. (\ref{eq:08}) while center manifold theory is performed to Eq. (\ref{eq:05}). Here $b_\pm=(-3\pm\sqrt{9+12\sqrt{-\lambda_1\lambda_2}/\alpha})/2$ and $c_\pm=(-3\pm\sqrt{9-12\sqrt{-\lambda_1\lambda_2}/\alpha})/2$, which give $b_+>0$, $b_-<0$ and ${\rm Re}(c_\pm)<0$.}
\label{tab:01}
\begin{tabular*}{\hsize}{@{}@{\extracolsep{\fill}}ccccccc@{}}
  \hline\hline
  Label & $(x_1,x_2,\lambda,\nu)$ & Existence & $\Omega_\phi$ & Eigenvalues & Stability & Method \\
  \hline
  O & $(0,0,\lambda,\nu)$ & All $\lambda_2$ & 0 & $[0,0,\frac{3(w_{\rm m}-1)}{2},\frac{3(1+w_{\rm m})}{2}]$ & saddle & linear stability theory \\
  A & $(0,1,0,0)$ & $\lambda_2=0$ & 1 & $[0,0,-3,-3(1+w_{\rm m})]$ & saddle & center manifold theory \\
  B & $(0,1,0,\sqrt{-6\lambda_1\lambda_2}/\alpha)$ & $\lambda_2<0$ & 1 & $[0,-3(1+w_{\rm m}),b_+,b_-]$ & saddle & linear stability theory \\
  C & $(0,1,0,-\sqrt{-6\lambda_1\lambda_2}/\alpha)$ & $\lambda_2<0$ & 1 & $[0,-3(1+w_{\rm m}),c_+,c_-]$ & stable & center manifold theory \\
  \hline\hline
\end{tabular*}
\end{table*}

Point O means the Universe is dominated by normal matters, which is unstable as we expected. Point C is the only stable attractor and stands for the cosmological solution where the Universe is dominated by scalar field with $w_{\phi}=-1$. This is inconsistent with the physical scenario we expected. Thus we would require $\lambda_2\geqslant0$ to avoid point C for the viable model. This requirement also makes the saddle point B disappear. Point A is a saddle point and we will discuss more about it later. The terrible thing is, unlike the results of most dark energy models (see Ref. \cite{Bahamonde2018} for a review), here the critical points do not provide any quantitative information about the evolution of the system for $\lambda_2>0$.

\subsection{The viable parameter space}
We want to find the allowed region in the parameter space in which the sine oEoS model could present a reasonable cosmological background evolution. Our discussion strongly depends on the results of the exponential potential. Here we summarize the main results obtained in Ref. \cite{Copeland1998}. For the potential $V(\phi)=V_0\exp(-\lambda\phi)$, where $\lambda$ is a constant, if $\lambda^2<3(1+w_{\rm m})$, then point $(x_1,x_2)=(\lambda/\sqrt{6},\sqrt{1-\lambda^2/6})$ is the only stable critical point and represents a Universe dominated by the scalar field with $w_\phi=-1+\lambda^2/3$. If $\lambda^2>3(1+w_{\rm m})$, then point $(x_1,x_2)=(\sqrt{3/2}(1+w_{\rm m})/\lambda,\sqrt{3(1-w_{\rm m}^2)/(2\lambda^2)})$ is the only stable critical point and represents a scaling solution with $\Omega_\phi=3(1+w_{\rm m})/\lambda^2$ and $w_\phi=w_{\rm m}$. Intuitively, in the sine oEoS model, we may can use a large $\lambda_1$ to realize $\rho_\phi$ tracks $\rho_{\rm m}$ and use a small $\lambda_2$ to accelerate the Universe.

\begin{figure*}[!t]
  \centering
  \includegraphics[width=0.99\textwidth]{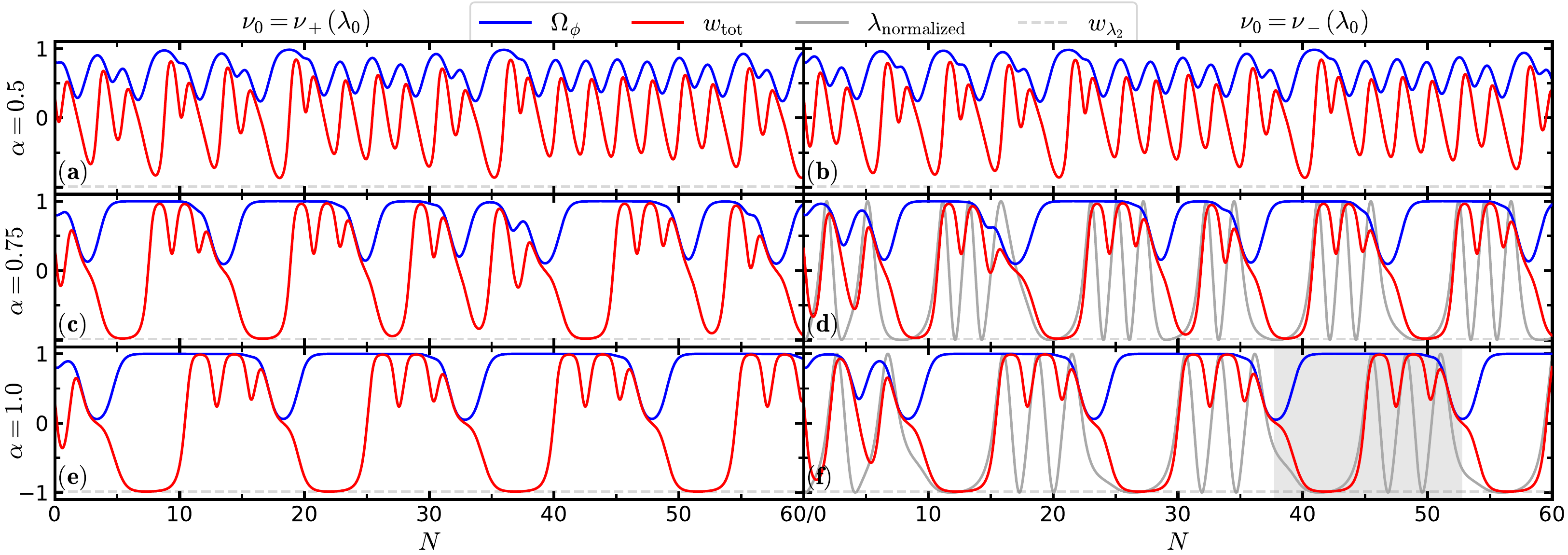}
  \caption{Evolution of the dark energy relative energy density $\Omega_\phi$ and the total effective EoS parameter $w_{\rm tot}$ for the sine oEoS model. The parameters are $w_{\rm m}=0$, $\lambda_1=4.5$, $\lambda_2=0.2$, $\alpha=0.5$, $0.75$ and $1.0$ for the first, second and third row, respectively. The initial conditions are $x_{1,0}=0.75$, $x_{2,0}=0.5$, $\lambda_0=0.3$, $\nu_0=\nu_+(\lambda_0)$ and $\nu_0=\nu_-(\lambda_0)$ for the first and second column, respectively. The plots start at $N=0$ and end at $N=60$. Here $w_{\lambda_2}=-1+\lambda_2^2/3$ and $\lambda_{\rm normalized}\equiv(2\lambda-\lambda_1-\lambda_2)/(\lambda_1-\lambda_2)=\cos(\phi/\alpha)$, which is plotted in the subplots (d) and (f) and can be used to track the position of $\phi$ in $V(\phi)$. Note that $\lambda_{\rm normalized}\approx1$ corresponds to $\lambda\approx\lambda_1$ and $\lambda_{\rm normalized}\approx-1$ corresponds to $\lambda\approx\lambda_2$.}
  \label{fig:02}
\end{figure*}

To solve the coincidence problem, we expect $\rho_\phi$ and $\rho_{\rm m}$ are in the same order of magnitude many times across the whole cosmic history. In the limit of $\alpha\ll1$, Eq. (\ref{eq:02}) can be approximately regarded as a single exponential potential with $\lambda=(\lambda_1+\lambda_2)/2$. In this limit, we expect $\rho_{\phi}$ tracks $\rho_{\rm m}$ in both radiation and matter era, which requires $(\lambda_1+\lambda_2)^2/4>{\rm max}(3[1+w_{\rm m}])=4$, i.e., $\lambda_1+\lambda_2>4$. Increasing $\alpha$ makes the scaling solution disappear and the ratio $\rho_\phi/\rho_{\rm m}$ time-dependent. However, numerical results show that generally $\lambda_1+\lambda_2>4$ is sufficient to satisfy the requirement that $\rho_\phi$ is in coincidence with $\rho_{\rm m}$ many times even for $\alpha=\mathcal{O}(1)$. In addition, very large $\alpha$ is not allowed because increasing $\alpha$ reduces the frequency of coincidence as shown in Fig. \ref{fig:02}. Comparing the left and right sides of Fig. \ref{fig:02}, we find the coincidence frequency is independent of the initial conditions. The exact upper limit on $\alpha$ may be subjective and a reasonable one can be $\alpha\lesssim\mathcal{O}(1)$.

\begin{figure}[!t]
  \centering
  \includegraphics[width=0.49\textwidth]{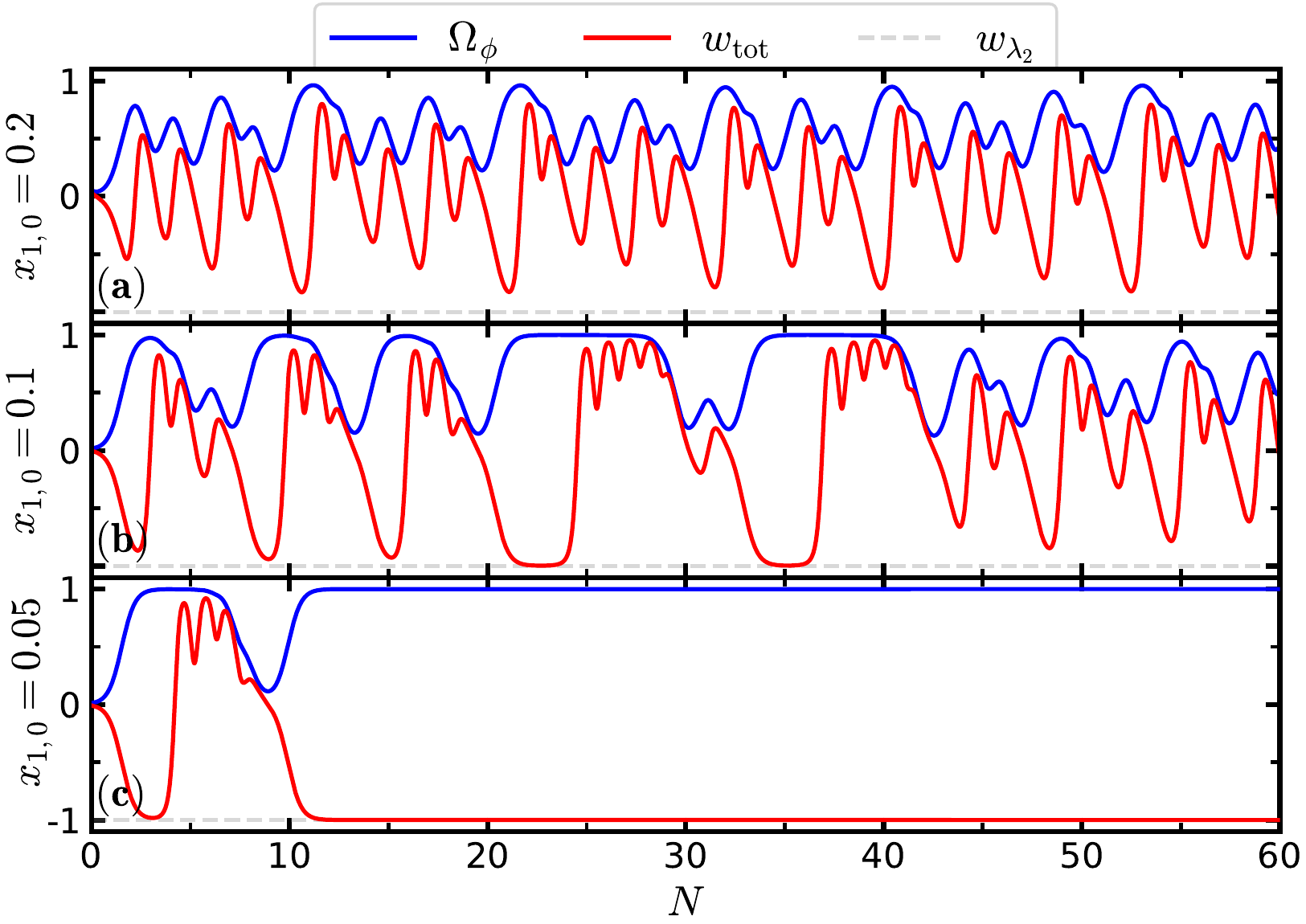}
  \caption{Evolution of $\Omega_\phi$ and $w_{\rm tot}$ for the sine oEoS model. The parameters are $w_{\rm m}=0$, $\lambda_1=5$, $\lambda_2=0$ and $\alpha=0.4$. The initial conditions are $x_{2,0}=0.1$, $\lambda_0=0$, $\nu_0=0$, $x_{1,0}=0.2$, $0.1$ and $0.05$ for the first, second and third row, respectively.}
  \label{fig:03}
\end{figure}

To explain the cosmic late-time acceleration, we need $w_\phi$ can be very close to $-1$ in some time period (see Eq. (51) in Ref. \cite{Aghanim2018} for observational constraints). In the limit of $\alpha\gg1$, locally we may can regard Eq. (\ref{eq:02}) as a single exponential potential. The minimum value of $w_\phi$ should be reached at $\lambda\approx\lambda_2$ and $w_{\phi,{\rm min}}\approx-1+\lambda_2^2/3$. This result is also numerically verified for $\alpha=\mathcal{O}(1)$. If we require $w_{\phi,{\rm min}}<-0.95$ as given in Ref. \cite{Aghanim2018}, then $\lambda_2<0.39$. In addition, very small $\alpha$ is not allowed because decreasing $\alpha$ increases the value of $w_{\phi,{\rm min}}$ (and also $w_{\rm tot,min}$ as shown in Fig. \ref{fig:02}). Unfortunately, the exact lower limit on $\alpha$ is not obtained here. One important thing is worth mentioning. For $\lambda_2=0$, the stability analysis summarized in Table \ref{tab:01} shows point A is a saddle point. Interestingly, this point can however attract many nontrivial solutions (see Fig. \ref{fig:03} for an example, which shows the scalar field with sufficient low kinetic energy will be trapped into point A). In order to improve the robustness of the sine oEoS model, it is reasonable to require $\lambda_2>0$. In summary, the viable parameter space should be $\lambda_1+\lambda_2>4$, $0<\lambda_2<0.39$, $\alpha=\mathcal{O}(1)$ and $V_0$ is arbitrary. If we assume $\phi=\mathcal{O}(1)$ at the onset of the cosmic Big Bang, it is reasonable to assume $V_0=\mathcal{O}(l_{\rm P}^{-2})$, where $l_{\rm P}$ is the Planck length. The sine oEoS model can explain the late-time acceleration with only one Planck scale parameter and several dimensionless parameters of order unity. In this sense no parameters need fine-tuning in our model.

In this paper, we do not perform complete parameter constraints with real data (see the next subsection for reasons), but we do find a set of parameters that make the sine oEoS model very close to the standard $\Lambda$CDM model in the late-time Universe. For example, one can easily verify $\Omega_{\rm m}=0.29$, $\Omega_\phi=0.71$, $w_{\rm tot}=-0.70$, and $\dx w_{\rm tot}/\dx N=-0.61$ at $N=59.28$ in Fig. \ref{fig:02} (d), where $\dx w_{\rm tot}/\dx N$ can be calculated based on Eq. (\ref{eq:08}) and the derivative of Eq. (\ref{eq:06}). In principle, we can set $N=59.28$ as today and set $N$ equals to a number smaller than zero as the beginning of the Big Bang if necessary. For the $\Lambda$CDM model, we know $\Omega_{\rm m}\approx0.3$, $\Omega_\Lambda\approx0.7$, $w_{\rm tot}\approx-0.7$ and $\dx w_{\rm tot}/\dx N\approx-0.63$ at today. Therefore, it is reasonable to believe that the sine oEoS model can well fit the observations about the late-time acceleration.

\subsection{Chaos}
Chaos appears in Fig. \ref{fig:02} and Fig. \ref{fig:03}. We think two phenomena are related to the emergence of chaos. One is that no critical point is stable for $\lambda_2\geqslant0$. The other is that the evolution of the scalar field is not attracted to the scaling solution when $\lambda\approx\lambda_1$ as shown in the shaded region in Fig. \ref{fig:02} (f). This is understandable since the only stable critical point for $V(\phi)\propto\exp(-\lambda_1\phi)$ is a spiral (see \cite{Copeland1998} and note that $\lambda_1$ is large enough in our model), not a node, and the $\lambda\approx\lambda_1$ part in the sine oEoS model is too short to successfully attract the scalar field. In contrast, the only stable critical point for $V(\phi)\propto\exp(-\lambda_2\phi)$ is a node \cite{Copeland1998} and can attract the scalar field faster. This result violates our initial idea that the Universe is decelerating when $\lambda\approx\lambda_1$ and accelerating when $\lambda\approx\lambda_2$. However, the model can still be used to realize the desired scenario and to solve the fine-tuning and coincidence problems (see discussions in Sec. \ref{sec:01} and the previous subsection).

The worse thing is that chaos make cosmological constraints tricky. In some classical dark energy models, the tracker property makes the cosmic late-time evolution independent of the dark energy initial conditions \cite{Steinhardt1999,Zlatev1999,Zhao2006b}. In this case, we do not need to consider these initial conditions in the cosmological constraints. However, in the sine oEoS model, the late-time evolution depends on the initial conditions of the scalar field. Furthermore, if we set $N=0$ as the beginning of the Big Bang and set $N=60$ as today, then the dependence should be quite strong. The consequence is that we have to consider the initial conditions as fitting parameters in the cosmological constraints and the resulting posterior distribution used in the classical statistical analysis changes dramatically with respect to these parameters. There should be many peaks in the posterior distribution, which makes the contour plots not reflect the parameter distributions correctly.

\section{Discussion}
In conclusion, qualitatively, the scenario described in Sec. \ref{sec:01} seems like a natural and simple way to eliminate both the fine-tuning and coincidence problems. Quantitatively, we have demonstrated the availability of the sine oEoS model proposed in this paper. However, our model seems to lead to some tricky consequences. The central problem is that the cosmic evolution strongly depends on the initial conditions in our model. On the one hand, this dependence leads to the technical difficulty in quantitatively constraining the model's parameters with observational data. On the other hand, the strong dependence in a sense is another fine-tuning problem if the allowed initial conditions are very rare. Figure \ref{fig:02} indicates that reasonable initial conditions and parameter settings do exist. However, whether these available initial conditions and parameter settings are widespread needs further exploration.

In addition to explaining the cosmic late-time acceleration, our model also provides an acceleration phase for the early Universe. This early acceleration can be used to solve horizon problem, which is related to the inflation theory \cite{Guth1981}. However, at least for now, we cannot conclude that our model provides a successful inflationary scenario. The reason is that many issues including the flatness problem and the initial conditions of perturbations have not been analyzed in this paper. Perhaps even worse, Eq. (\ref{eq:02}) involves an exponential potential with an extra sine function and current observational data disfavor the exponential type of inflation potentials \cite{Martin2014}. This do not directly disable our model in the early Universe and further analysis is needed. Work is currently underway to provide a quantitative analysis of the inflationary stage for our model.

Over the past forty years, cosmologists have been convinced that the Universe has gone through two acceleration phases: inflation (the early-time acceleration) and the late-time acceleration. In Sec. \ref{sec:01}, we pointed out that multi-acceleration scenario may be the key to solving the cosmological coincidence problem. There are other clues suggesting that the middle-time cosmic evolution may differ from the predictions of the standard cosmological model, e.g., the early dark energy scenario recently proposed to ease the Hubble tension \cite{Karwal2016,Mortsell2018,Poulin2019,Sakstein2019}. We are publishing this paper in the hope that it will highlight the problem: Is there acceleration phase in the middle-time Universe?

\section*{Acknowledgements}
This work was supported by the National Natural Science Foundation of China under Grant Nos. 11633001 and 11920101003, and the Strategic Priority Research Program of the Chinese Academy of Sciences, Grant No. XDB23000000.

\appendix
\section{Comments on the existing oscillating dark energy models}\label{sec:A}
Here we discuss whether the existing oscillating dark energy models \cite{Xia2005,Xia2006,Liu2009,Nojiri2006a,Nojiri2006b,Kurek2010,Lazkoz2010,Pan2018,Panotopoulos2018, Brownsberger2019,Rezaei2019,Tamayo2019} can realize the physical scenario that the Universe is dominated by normal matters and dark energy alternately across the whole cosmic history. Solving the energy conservation equation $\dot{\rho}+3(1+w)H\rho=0$, we obtain
\begin{equation}\label{eq:A1}
  \rho\propto\exp(\int_0^z\frac{3[1+w(\tilde{z})]}{1+\tilde{z}}\dx\tilde{z})
  =\exp[3\int_0^N1+w(\tilde{N})\dx\tilde{N}],
\end{equation}
where $z$ is the redshift, $N\equiv\ln(a_0/a)$ and $a_0$ is the value of the scale factor at today (note that here the definition of $N$ is different from that in the main text). For the normal matters, Eq. (\ref{eq:A1}) gives the dust density $\rho_{\rm dust}\propto(1+z)^3$ and the radiation density $\rho_{\rm r}\propto(1+z)^4$. For the dark energy with $\sin z$-type oscillation $w_{\rm DE}=w_0+w_1\sin z$ \cite{Xia2005}, Eq. (\ref{eq:A1}) gives $\rho_{\rm DE}\propto(1+z)^{3+3w_0}$ for high redshift (note that $\lim_{z\rightarrow+\infty}\int_0^z\frac{\sin\tilde{z}}{1+\tilde{z}}\dx\tilde{z}={\rm const.}$). In this case, on the one hand, $\rho_{\rm DE}$ cannot track the normal matter density in both the early and late-time Universe. On the other hand, no deceleration-acceleration transition occurs in the early Universe. Thus this model cannot realize the desired scenario. For the dark energy with $\sin N$-type oscillation $w=w_0+w_1\sin(w_2N+w_3)$ \cite{Xia2006,Liu2009}, the model satisfies our requirement if there is only one type of normal matter in the Universe. However, the realistic Universe contains radiation and dust. Thus this model is also invalid. Similar discussions also apply to the other type of EoS discussed in Refs. \cite{Nojiri2006a,Nojiri2006b,Kurek2010,Lazkoz2010,Pan2018,Panotopoulos2018,Brownsberger2019,Rezaei2019,Tamayo2019}.


%

\end{document}